\documentstyle[preprint,aps,epsf]{revtex}
\def \Et {E_T}

\newcommand{\MET}{\mbox{$\protect \raisebox{.3ex}{$\not$}\Et$}}

\def\Z0{${\em Z^0\/}$}

\def\r#1 {$^{#1}$}

\newcommand{\ttbar}{t\bar{t}}

\begin{document}
\draft
\title{
\begin{flushright}
{\normalsize\rm
FERMILAB-PUB-98/319-E  }
%CDF/ANAL/TOP/PUBLIC/4619 \\
% \today}
\end{flushright}
 Measurement of the Top Quark Mass with 
 the Collider  Detector  at Fermilab }
\author{
%From:	FNALD::CAROL        23-SEP-1998 06:00:00.55
%To:	FNALD::GALTIERI
%CC:	
%Subj:	CDF Default Author List!
%
\font\eightit=cmti8
\def\r#1{\ignorespaces $^{#1}$}
\hfilneg
\begin{sloppypar}
\noindent
F.~Abe,\r {17} H.~Akimoto,\r {39}
A.~Akopian,\r {31} M.~G.~Albrow,\r 7 A.~Amadon,\r 5 S.~R.~Amendolia,\r {27} 
D.~Amidei,\r {20} J.~Antos,\r {33} S.~Aota,\r {37}
G.~Apollinari,\r {31}
% ...
\end{sloppypar}
\vskip .026in
\begin{center}
(CDF Collaboration)
\end{center}

\vskip .026in
\begin{center}
\r 1  {\eightit Argonne National Laboratory, Argonne, Illinois 60439} \\
\r 2  {\eightit Istituto Nazionale di Fisica Nucleare, University of Bologna,
I-40127 Bologna, Italy} \\
\r 3  {\eightit Brandeis University, Waltham, Massachusetts 02254} \\
\r 4  {\eightit University of California at Los Angeles, Los 
Angeles, California  90024} \\  
\r 5  {\eightit University of Chicago, Chicago, Illinois 60637} \\
\r 6  {\eightit Duke University, Durham, North Carolina  27708} \\
\r 7  {\eightit Fermi National Accelerator Laboratory, Batavia, Illinois 
60510} \\
\r 8  {\eightit University of Florida, Gainesville, Florida  32611} \\
\r 9  {\eightit Laboratori Nazionali di Frascati, Istituto Nazionale di Fisica
               Nucleare, I-00044 Frascati, Italy} \\
\r {10} {\eightit University of Geneva, CH-1211 Geneva 4, Switzerland} \\
\r {11} {\eightit Harvard University, Cambridge, Massachusetts 02138} \\
\r {12} {\eightit Hiroshima University, Higashi-Hiroshima 724, Japan} \\
\r {13} {\eightit University of Illinois, Urbana, Illinois 61801} \\
\r {14} {\eightit Institute of Particle Physics, McGill University, Montreal 
H3A 2T8, and University of Toronto,\\ Toronto M5S 1A7, Canada} \\
\r {15} {\eightit The Johns Hopkins University, Baltimore, Maryland 21218} \\
\r {16} {\eightit Institut f\"{u}r Experimentelle Kernphysik, 
Universit\"{a}t Karlsruhe, 76128 Karlsruhe, Germany} \\
\r {17} {\eightit National Laboratory for High Energy Physics (KEK), Tsukuba, 
Ibaraki 305, Japan} \\
\r {18} {\eightit Ernest Orlando Lawrence Berkeley National Laboratory, 
Berkeley, California 94720} \\
\r {19} {\eightit Massachusetts Institute of Technology, Cambridge,
Massachusetts  02139} \\   
\r {20} {\eightit University of Michigan, Ann Arbor, Michigan 48109} \\
\r {21} {\eightit Michigan State University, East Lansing, Michigan  48824} \\
\r {22} {\eightit University of New Mexico, Albuquerque, New Mexico 87131} \\
\r {23} {\eightit The Ohio State University, Columbus, Ohio  43210} \\
\r {24} {\eightit Osaka City University, Osaka 588, Japan} \\
\r {25} {\eightit Universita di Padova, Istituto Nazionale di Fisica 
          Nucleare, Sezione di Padova, I-35131 Padova, Italy} \\
\r {26} {\eightit University of Pennsylvania, Philadelphia, 
        Pennsylvania 19104} \\   
\r {27} {\eightit Istituto Nazionale di Fisica Nucleare, University and Scuola
               Normale Superiore of Pisa, I-56100 Pisa, Italy} \\
\r {28} {\eightit University of Pittsburgh, Pittsburgh, Pennsylvania 15260} \\
\r {29} {\eightit Purdue University, West Lafayette, Indiana 47907} \\
\r {30} {\eightit University of Rochester, Rochester, New York 14627} \\
\r {31} {\eightit Rockefeller University, New York, New York 10021} \\
\r {32} {\eightit Rutgers University, Piscataway, New Jersey 08855} \\
\r {33} {\eightit Academia Sinica, Taipei, Taiwan 11530, Republic of China} \\
\r {34} {\eightit Texas A\&M University, College Station, Texas 77843} \\
\r {35} {\eightit Texas Tech University, Lubbock, Texas 79409} \\
\r {36} {\eightit Istituto Nazionale di Fisica Nucleare, University of Trieste/
Udine, Italy} \\
\r {37} {\eightit University of Tsukuba, Tsukuba, Ibaraki 315, Japan} \\
\r {38} {\eightit Tufts University, Medford, Massachusetts 02155} \\
\r {39} {\eightit Waseda University, Tokyo 169, Japan} \\
\r {40} {\eightit University of Wisconsin, Madison, Wisconsin 53706} \\
\r {41} {\eightit Yale University, New Haven, Connecticut 06520} \\
\end{center}

}
\date {\today }
\maketitle 
\begin{abstract}      
 We present a new measurement of the top quark mass in $t\bar t$ events
 in which both $W$ bosons from top quarks decay into leptons 
($e \nu$,$\mu \nu$).
 We use events collected by the CDF experiment from
 $p\bar p$ collisions at $\sqrt s$=1.8 TeV at the Tevatron
 collider.   We 
 measure a top 
quark mass of 167.4~$\pm$~10.3(stat)~$\pm$~4.8(syst)~GeV/$c^2 $ from
a sample of eight events. We combine
this result with previous CDF measurements in other decay channels to
obtain a mass value of 176.0~$\pm$~6.5~GeV/$c^2$.
\end{abstract}
\pacs{PACS numbers: 14.65.Ha, 13.85.Qk, 13.85.Ni }
% REVTEX galley needs narrowtex
%\narrowtext
%
%%************************************************************************
%% Introduction 
%%************************************************************************

 A precise measurement of the top quark mass is an important ingredient
 in testing the consistency of the standard model with experimental
 data. In addition,  precise $W$ and top mass measurements can 
provide information
 on the mass of the Higgs boson, which is a remnant of the mechanism that
 gives rise to spontaneous electroweak
 symmetry breaking. 
Using 109 pb$^{-1}$ of
 data accumulated by the CDF experiment at the Fermilab Tevatron from 1992 
 through 1995,
 we report an improved measurement of the top quark mass using 
 dilepton events originating predominantly from  
 $t\bar t \rightarrow W^+b W^-\bar b \rightarrow 
(\ell^+\nu b)(\ell^-\bar \nu \bar b)$,
 where $\ell = e$ or $ \mu$. This measurement supersedes our previously
 reported  result in the dilepton channel~\cite{cdf_dilmass}. The
 previous result was obtained 
 by comparing data with Monte Carlo simulation of $\ttbar$ events
 for two kinematic variables, the $b$-jet energies and the 
invariant masses of the lepton and $b$-jet systems.
Here we make use of all available information in the event and obtain
a more precise measurement.
We combine the result from the dilepton channel with those from the
lepton plus jets channel
($t\bar t \rightarrow W^+b W^-\bar b \rightarrow 
(\ell^+\nu b)(q \bar q' \bar b)$) ~\cite{ljet} and the 
all-hadronic channel
($t\bar t  \rightarrow 
(q \bar q' b)(\bar q q' \bar b)$)~\cite{allh} 
to obtain an overall top quark mass value from the CDF data.

%%***********************************************************************
%% detector description 
%%************************************************************************

The CDF detector consists of a magnetic spectrometer surrounded by calorimeters
and muon chambers~\cite{NIM}.  A four-layer silicon vertex detector
(SVX)~\cite{svx}, located immediately 
outside the beampipe, is surrounded by
the central tracking chamber (CTC) which is inside a 1.4 T superconducting
solenoid. This tracking system is used to measure the momenta 
of charged particles.
 Electromagnetic and hadronic calorimeters, located outside the CTC, 
are segmented in projective towers and cover the
pseudorapidity region $|\eta| $ $<$ 4.2~\cite{coord}.
They are used to identify electron and photon candidates and jets, and 
are used to measure the missing
transverse energy ($\MET$) which can indicate the presence of
energetic neutrinos.  Outside the calorimeters, drift chambers in the region
$|\eta|$ $ <$ 1.0 provide for muon identification.  
A three-level trigger selects events that contain a high $P_T$ electron 
or muon
for this analysis. 

%%*******************************************************************
%% sample selection: need to expand 
%%*******************************************************************

   We apply the same event selection criteria and use identical background
 calculations 
 as those employed in the previous mass 
 analyses of the dilepton channel~\cite{cdf_dilmass}. We 
require two high transverse momentum ($P_T >$ 20
GeV/c) oppositely charged leptons ($e$ or $\mu$) in the central detector
region ($|\eta| <$ 1), with at least one of them well isolated from
nearby tracks and calorimeter activity. To reject 
 $Z$  $\rightarrow \ell^+ \ell^- X$ events we require that 
the dilepton 
invariant mass, $M_{e e}$ or $M_{\mu \mu}$, be outside the interval
75-105 GeV/$c^2$, and remove events containing an isolated photon
with $E_T > 10 $ GeV if they are consistent with radiative $Z$ decays.
 We require at least two jets in the region
 $|\eta| <$ 2.0, each with observed $\Et > $ 10 GeV in a cone radius 
$\Delta R=\sqrt{\Delta \eta^2+\Delta \phi^2}=0.4$, and require
$|\MET| >$ 25 GeV as a signature for missing neutrinos. 
To reject events in which $\MET$ is due to lepton or jet energy mismeasurements
we require $|\MET| >$ 50 GeV if $\MET$ is close 
 to a lepton or a jet ($\Delta \phi (\MET, \ell$~or~$j) < 20^{\circ}$).
Finally, we require $H_T > 170 $ GeV, where 
 $H_T$ is the scalar sum of the $|P_T|$ of the two leptons ($E_T$ for
 electrons), the $E_T$ of the two highest $E_T$ jets and $|\MET|$.
 We obtain a sample of eight candidate events. The expected 
 background of $1.3\pm 0.3$ events 
 consists of 
 events in which a track or a jet is misidentified as
 a lepton (0.29 events), 
 Drell-Yan production (0.35 events), $WW$ production (0.24 events),  
 $Z\rightarrow \tau \tau$ decays (0.26 events) and $Z\rightarrow \mu \mu$ 
 decays in which $\mu$ tracks are mismeasured (0.20 events). 

To extract a top mass measurement from these events 
the jet energies are corrected for losses in cracks between detector 
components, absolute energy scale, contributions from the underlying event
and multiple interactions, and losses outside the clustering cone. These 
corrections are determined from a 
combination of Monte Carlo simulation and data~\cite{ljet}. 
An additional energy correction is applied to jets containing
a muon to take into account the low calorimeter response to muons (we add
the $E_T$ of the muon to the energy)
and missing neutrinos from semileptonic decays of heavy quarks in the jets. We
obtain a corrected $\MET$ by correcting each of its components, $\it{i.e.}$,
the jets, the leptons and the energy which has not been clustered into
jets.

%%*******************************************************************
%% technique  and procedures 
%%*******************************************************************
The procedure used here has two steps: we reconstruct each event to
obtain a top mass estimate for the event, then we apply a 
likelihood method to obtain
an overall mass value in the presence of background.
Each candidate event is reconstructed according to the $t\bar t$ decay 
hypothesis in the dilepton channel: 
\begin{eqnarray*}
  t \rightarrow W^+ b \rightarrow \ell_1^+\nu_1 b \\
  \bar t \rightarrow W^- \bar b \rightarrow \ell_2^- \bar \nu_2 \bar b
\end{eqnarray*}
The two highest $E_T$ jets in the event are assumed to be the $b$-jets
from top decays. We assume the $b$-jet mass to be 5 GeV/$c^2$.
After applying the invariant mass constraints $m(\ell_1\nu_1) = 
m(\ell_2 \bar \nu_2)= m_W$ and 
$m(\ell_1\nu_1 b ) = m(\ell_2 \bar \nu_2 \bar b ) $, 
the system remains underconstrained due to the two unmeasured neutrinos.
Therefore, for any assumed top mass value $m_t$, 
we use a weighting technique to determine a function, $f(m_t)$,
from which we extract a top mass value~\cite{dg-d0}. We proceed as follows.
We assume a top quark mass ($m_{t}$) 
and the two neutrino $\eta$ values ($\eta_1$, $\eta_2$, see below) and
solve for the neutrino momenta, up to a four-fold 
ambiguity (two $P_z(\nu)$ choices for each $\nu$)
for each of the two jet charged-lepton pairings.
We then assign a weight to each solution 
by comparing $\MET^p$, the sum of the neutrino transverse momenta
for that solution, to $\MET^m$, the measured missing 
transverse energy after proper correction:
\begin{eqnarray*}
 g(m_t,\eta_1,\eta_2) = exp{\left( -\frac{(\MET^p - \MET^m)^2} {2\sigma^2}
                          \right)}  
\end{eqnarray*}                
 where $\sigma$ is
the $\MET$ resolution for that event (see below). 

For each choice of $m_t,~\eta_1,$ and~$\eta_2 $
we take into account the detector 
 resolution for jets and leptons by 
sampling ($\it{i.e.}$ fluctuating) the measured quantities many times 
according to their resolutions. 
The Gaussian resolutions for electrons and muons are 
$\sigma_E/E = \sqrt{(0.135)^2/E_T+(0.02)^2}$
and $\sigma(1/P_T)=0.11\%$, where $E_T$, in GeV, is the 
electron energy measured in the 
calorimeter and $P_T$ is the beam-constrained muon momentum measured in the CTC.
For jets we use an $E_T$ dependent 
 Gaussian resolution appropriate for $b$-partons derived from the
HERWIG $t\bar t$ Monte Carlo, in conjunction with detector simulation,
assuming a top mass of 170 GeV/$c^2$ ~\cite{topprd}.
The $\MET$ is recomputed for each sampling using the new
jet and lepton energies.  This procedure takes into account 
all the uncertainties in the $\MET$ measurement except for the 
resolution of the
unclustered energy measurement, which is estimated to be 4 GeV for
both transverse components for low luminosity events. Hence 
we use $\sigma$= 4$\sqrt{n}$ GeV in the 
above expression for $g(m_t,\eta_1,\eta_2)$, where $n$ is the number of
interactions in the event.
This value has been obtained from minimum bias events and for each event 
is properly scaled to take into account the effect of
multiple interactions at high luminosity.

For each assumed top mass value we use 
several (100) values of the two neutrino $\eta_1$ and $\eta_2$, chosen
from distributions obtained from
the HERWIG Monte Carlo predictions\cite{herwig}. They are consistent with
independent Gaussian distributions with $\sigma$ = 1.0 unit of $\eta$.
The weight is summed for all samplings as well as over all $\eta_1$,
$\eta_2$ values and all the eight possible combinations; thus for each event 
at each top mass, $m_t$,  we evaluate an overall weight:
\begin{eqnarray*}
  f(m_t) = \sum_{\eta_1,\eta_2,E_{T_i}, \ell_1, \ell_2} 
                    g(m_t,\eta_1,\eta_2)  
\end{eqnarray*}
where $E_{T_i}$  refers to all the jets in the event. 
We then compute the weight as a function
of the top mass in 2.5 GeV/$c^2$ steps in the range 90-290 GeV/$c^2$.

\begin{figure}
% galley size is 3.4
\epsfxsize=6.5in
\epsffile[50 150 540 640]{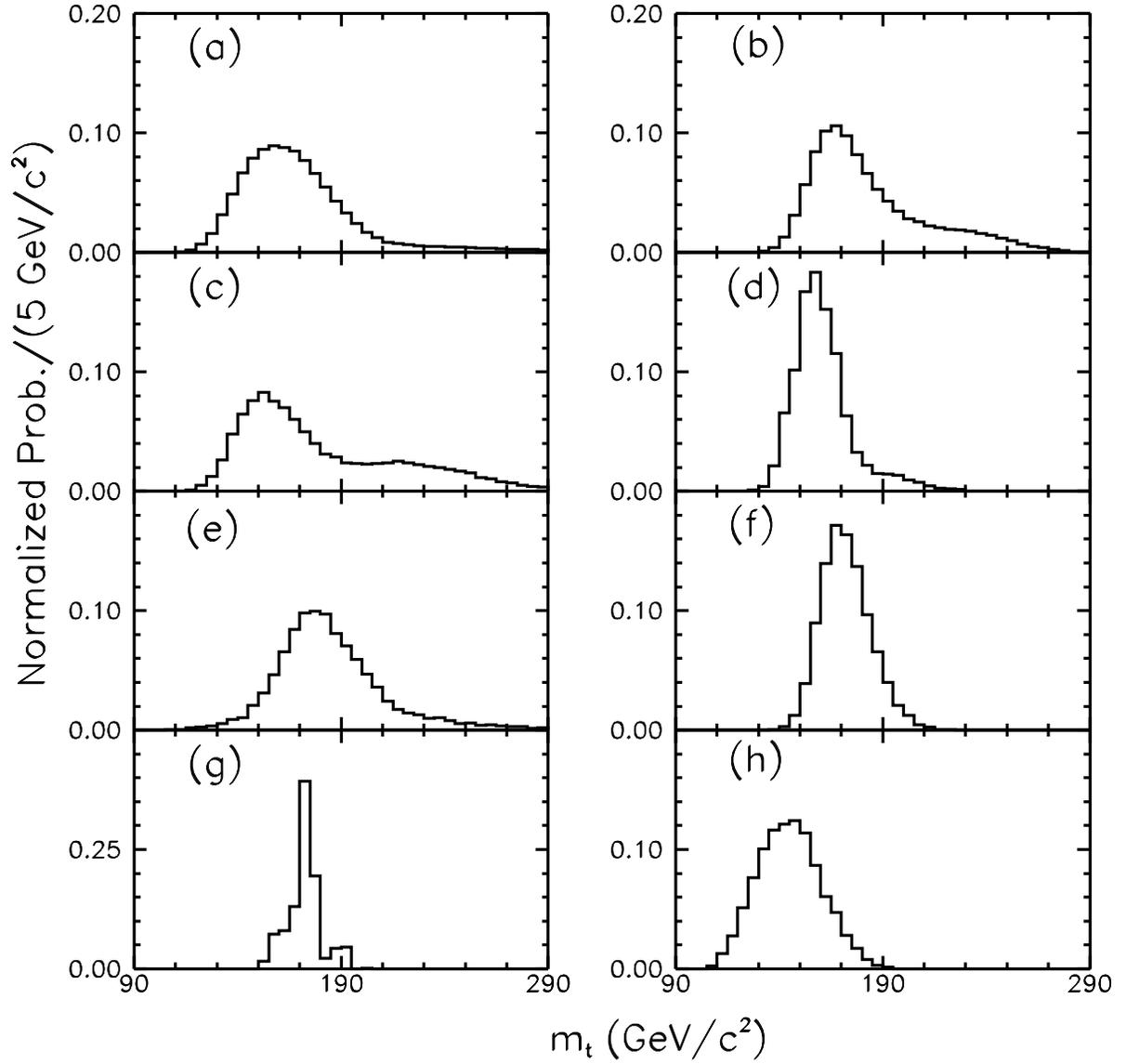}
\caption{Weight distribution normalized to unity as a function of 
        $m_t$  for the eight dilepton top candidate events (a-h).}
\label{weight}
\end{figure}

\begin{figure}
% REVTEX galley size is 3.4
\epsfxsize=6.5in
\epsffile[70 150 540 640]{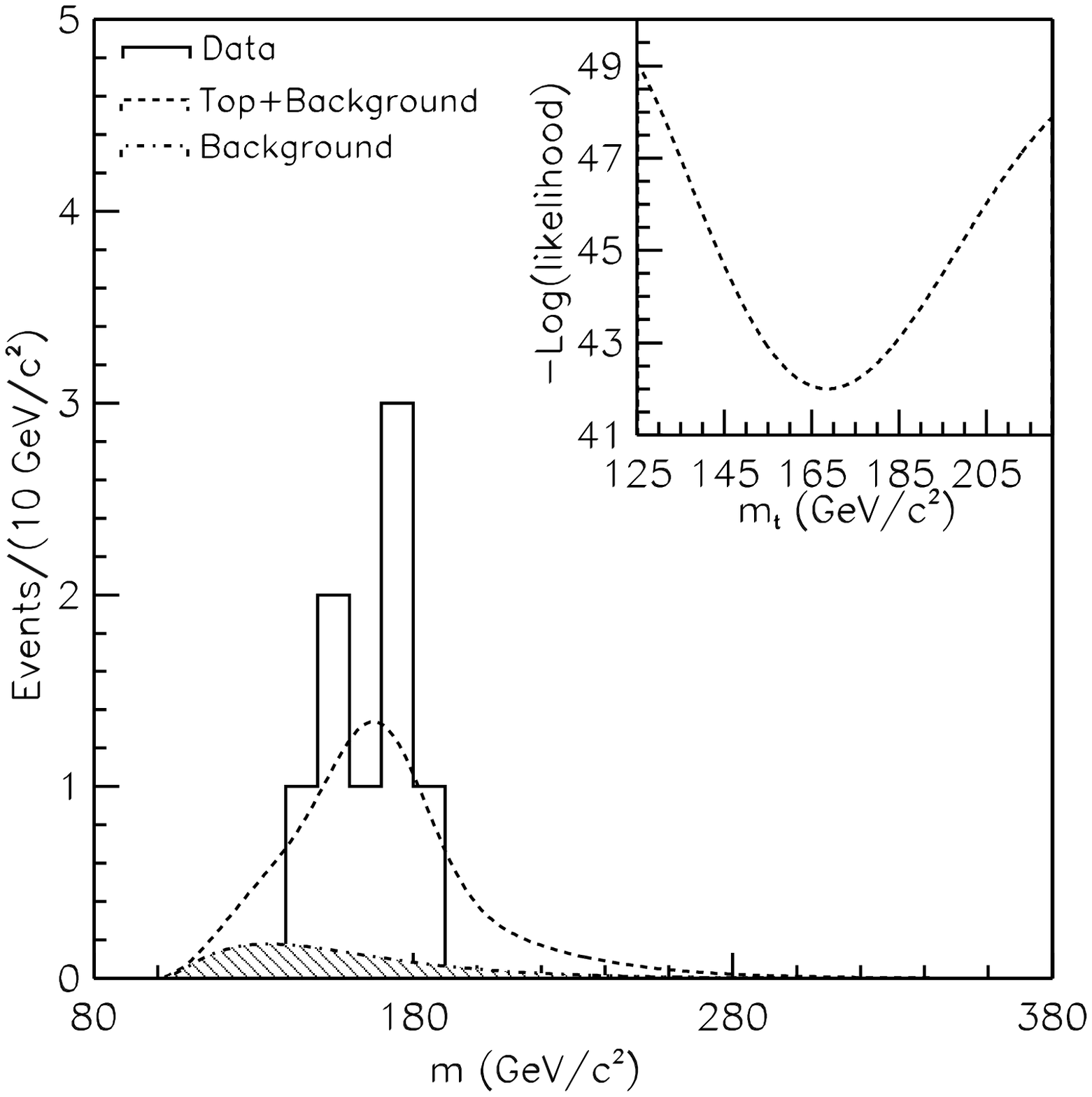}
\caption{Reconstructed top mass for the eight dilepton events (solid).
         Background distribution (shaded, 1.3 events) and 
         top Monte Carlo (6.7
         events) added to background (dashed).
         The negative log-likelihood distribution as a function of 
         the top mass is shown in the inset. }  
\label{mass}
\end{figure}

The $f(m_t)$ distribution for each of
the eight candidate events, normalized to unity, 
is shown in Figure~\ref{weight}. 
For each event, $i$, we use this 
distribution to determine a top mass estimate, $m_i$,
by averaging the values of $m_t$ corresponding 
to values of $f(m_t)$ closest to 
and greater than $f(m_t)_{max}$/2 on either side 
of the maximum.
 The $m_i$ distribution for the eight events is shown in Figure~\ref{mass},
 together with the Monte Carlo expectation for background alone,
 and top 
 plus background normalized to the data.

%%**********************************************************************
%% likelihood
%%*********************************************************************
  Given the $m_i$ distribution for a sample of $N$ events, 
we use  a maximum likelihood method 
similar to that employed in the lepton plus 
jets mass analysis\cite{topprd} to extract a top mass value.
 We write the likelihood as:
$${\cal L} = G(n_b;\bar n_b,\sigma_b) \cdot P(N;n_s+n_b) \cdot
  \prod_i {{n_s T(m_i,m_t) + 
      n_b B(m_i)} \over{ n_s + n_b}} $$
where $T(m_i,m_t)$ and $B(m_i)$ are the probability density functions 
(templates) for reconstructing a mass $m_i$ from $t\bar t$ 
events with a true top mass 
$m_t$ and from background events, respectively;
$n_s$ and $n_b$ are the numbers of $t\bar t$ and background events;
$G(n_b; \bar n_b, \sigma_b)$ is a Gaussian in $n_b$ with mean 
$\bar n_b$ (1.3 events) and width $\sigma_b$ (0.3 events)
 and $P(N;n_s+n_b)$ is a
Poisson distribution in N with mean $n_s+n_b$. 
  The top templates, $T(m_i,m_t)$, are obtained by performing
the dilepton mass reconstruction algorithm on HERWIG $t\bar t$ 
Monte Carlo samples, and  parameterizing the resulting
reconstructed mass distributions as 
 smooth functions of both $m_i$ and $m_t$\cite{ljet}.  
The background templates, $B(m_i)$, are derived 
from data and Monte Carlo~\cite{cdf_dilmass}. 
The likelihood fit gives the top mass measurement and a statistical
error taken from the change in the -log($\cal L$) of 0.5 units from its
minimum value.

  To check the likelihood fitting technique, we use Monte Carlo events
to perform many 
pseudo-experiments for several input top mass values. At each top mass 
value each experiment consists of a total of eight 
top and background events, with the number of background events
drawn from a binomial
distribution with a mean  of $1.3 \pm 0.3$ events. 
The mass estimate for each event is drawn from the appropriate
parameterized template. 
For each experiment
we obtain a mass, $m_{exp}$, and a statistical error, $\sigma_{exp}$.
We study the distributions of the pulls, $(m_{exp}-m_t)/\sigma_{exp}$,
and find that the medians are consistent with zero and the 
widths are at most 1.1.
We take this width into account by multiplying the error returned by the fit
by a factor 1.1. This factor is included in all statistical errors on the
dilepton measurements given below.

 We apply the likelihood method to the data shown in Figure~\ref{mass}.
  The inset shows the negative 
 log-likelihood as a function of the top mass, from which we determine a top 
 quark mass value of 167.4~$\pm$~10.3 (stat)~GeV/$c^2 $.   
  Monte Carlo studies performed
 with pseudo-experiments yield an 8\% probability for one such
 experiment to
 have a statistical error $\le 10.3$ GeV/$c^2$ at a mass of 167 GeV/$c^2$.
In comparison, the previous result was 161~$\pm$~17 (stat) 
GeV/$c^2$~\cite{cdf_dilmass}.
%********************************************************************
%*Cross checks: two svx tags  
%********************************************************************

 The mass reconstruction procedure used here is quite different
 from that used 
for the lepton plus jets sample~\cite{ljet}. In that case
there is only one missing neutrino, and a 
  kinematic fit with two constraints can be performed. We 
   cross-check the present procedure by applying it 
 to the five events of the 
lepton plus four jets mass sample in which two jets are tagged as $b$-jets 
by the SVX~\cite{ljet} . We assume that  the two 
 untagged jets are the products  of the hadronic $W$ decay and 
 mimic the dilepton decay by treating the highest $\Et$ untagged  
 jet as a lepton and the second untagged jet as a neutrino. 
  The present likelihood method then gives
a top quark mass from the five events of 181.5~$\pm$~12.6~GeV/$c^2$, which
differs by 11 GeV/$c^2$
from that obtained with the lepton plus jets kinematic fit procedures,
 170.1~$\pm$~9.3~GeV/$c^2$~\cite{ljet}. In comparison,
Monte Carlo studies show that the difference in mass obtained with
the two methods is expected to be centered at zero with a resolution of 
 14 GeV/$c^2$.

%********************************************************************
% measurements 
%********************************************************************

%%*******************************************************************
%% Systematic errors
%%********************************************************************
  The systematic errors on the mass measurement
are estimated with the same general procedure 
as in the lepton plus jets mass analysis~\cite{ljet}:
 we generate new top 
 mass distributions by varying the appropriate quantities in the Monte
 Carlo simulation and then performing likelihood fits to many experiments of
 eight events each using the standard templates.
 The mass shifts obtained determine the systematic errors, which
 are summarized in Table~\ref{sys}.
The largest contribution comes from the systematic 
uncertainties on the jet energy.
The second largest source, the uncertainty in modeling 
initial or final state radiation in $t\bar t$ events,
is estimated using events generated with
the PYTHIA Monte Carlo program\cite{pythia} to isolate the 
effects on the top mass due to initial and final state radiation jets.  
The error due to background shape is 
 estimated by refitting the data with three different backgrounds:
 $W^+W^-$, fake leptons, and the sum of the two with the
 appropriate weights. 
That due to the choice of parton distribution functions (0.6 GeV/$c^2$) 
is estimated by using an alternative set of functions 
in the HERWIG Monte Carlo.
The effect due to the choice of Monte Carlo generator (0.6 GeV/$c^2$)
is estimated by comparing the mass value obtained with events from the
HERWIG and PYTHIA Monte Carlo samples.
 The total systematic error amounts to 4.8 GeV/$c^2$.

%%********************************************************************
%% Conclusions 
%%********************************************************************

Table~\ref{sys} includes the systematic errors for the other two decay 
channels. 
 The systematic error due to hard gluon
  radiation uncertainty in the lepton plus jets~\cite{ljet}
  and the all-hadronic~\cite{allh} channels are
  now evaluated in the same way as described above. For the all-hadronic 
 case the old value of 8.0 GeV/$c^2$ was an overly conservative estimate.
 Also, the ``fit procedure'' systematic in the all-hadronic channel 
 (5.0 GeV/$c^2$)
 has been removed as being overly conservative. 
 The overall systematic error  for the lepton plus jets channel has increased
 from 4.9 to 5.3 GeV/$c^2$, and that for the 
 all-hadronic channel is reduced from  12.0 to 5.7 GeV/$c^2$.
 The new top mass measurement for the dilepton channel and those for the 
other two channels with revised systematic errors are shown 
 in Table~\ref{comb}. 
  
      The results for the three channels are combined
with standard methods~\cite{PDG} to yield
  an overall CDF mass measurement. The 
three statistical errors are taken
  as uncorrelated, while the systematic errors are assumed to be either
 entirely correlated or uncorrelated between any two channels. The
  primary systematic error, that due to jet energy 
 uncertainty, is
  taken as entirely correlated among all channels, as is the
  systematic error due to the Monte Carlo model used 
  (gluon radiation and simulation entries in Table~\ref{sys}).
      The combined result is 
\begin{center}
	$m_t$ = 176.0~$\pm$~6.5~GeV/$c^2$
\end{center}
including both statistical
  and systematic errors. 
In Table~\ref{comb} we show the combined value with separate statistical 
and systematic
errors. They are obtained by defining the combined statistical 
error as the sum in quadrature of the weighted individual statistical errors,
and the systematic error as the difference in quadrature of the total and 
statistical errors. 
The relative contributions from the three channels
  are 67\% for lepton plus jets, 18\% for dileptons and
 15\% for all-hadronic.

\vspace{0.1in}
     We thank the Fermilab staff and the technical staffs of the participating
institutions for their contributions.   This work was supported by the U.S.
Department of Energy and National Science Foundation, the Italian Istituto
Nazionale di Fisica Nucleare, the Ministry of Science, Culture, and Education of
Japan, the Natural Sciences and Engineering Research Council of Canada, 
the National Science Council of the Republic of China, and the
A. P. Sloan Foundation.

%\newpage

\newpage
\begin{table} [htbp]
\caption{\label{sys}Systematic errors (in Gev/$c^2$) on the top mass 
          measurement for the three $\ttbar$ decay channels.
          The Monte Carlo modeling term includes effects from: 
          parton density functions and b-tag uncertainty where
          applicable. For the lepton plus jet channel the Monte Carlo
          statistics term was included in the statistical error.    }
\begin{center}
\begin{tabular}{lccc}  
 Channel & dilepton & $\ell$+jets & all-had. \\  \hline
 Jet energy scale                  & 3.8  & 4.4  &  5.0  \\
 Initial and final state radiation & 2.7  & 2.6   &  1.8  \\
 Monte Carlo modeling              & 0.6  & 0.5   &  0.2  \\
 Monte Carlo generator             & 0.6  & 0.1   &  0.8  \\
 Background shape                  & 0.3  & 1.3   &  1.7   \\ 
 Monte Carlo statistics            & 0.7  & n.a.  &  0.6  \\  \hline
  Total                            & 4.8  & 5.3   &  5.7  \\
\end{tabular}
\end{center} 
\end{table}

\begin{table}[htbp]
\caption{\label{comb} Summary of top mass measurements with the CDF detector}
\begin{center}
\begin{tabular}{lcc} 
 Channel       & Top mass (GeV/$c^2) $   & Reference \\  \hline
 Dilepton      &167.4~$\pm$~10.3 $\pm$~4.8 & this paper  \\
 Lepton + jets &175.9~$\pm$~4.8 $\pm$~5.3  & ~\cite{ljet}, this paper  \\
 All-hadronic  &186.0~$\pm$~10.0~$\pm$~5.7 & ~\cite{allh}, this paper \\  \hline
 Combined      &176.0~$\pm$~4.0 $\pm$~5.1  & this paper  \\
\end{tabular}
\end{center}
\end{table}

%\clearpage
 
%\pagestyle{empty}



\begin{thebibliography}{99}
\bibitem{coord} In the CDF coordinate system, $\theta$ and $\phi$ are the 
 polar and azimuthal angles, respectively, with respect to the proton beam 
 direction which defines the $z$ axis. The pseudorapidity $\eta$ is defined as 
 $-\ln(\tan\frac{\theta}{2})$. The transverse momentum of a particle with 
 momentum $P$ is 
 $P_T =  P \sin\theta$. The analogous quantity using energy, defined as 
 $\Et =  E \sin\theta$, is called transverse energy. 
 The missing transverse energy $\MET$ is defined as 
 \mbox{$-\Sigma \Et^i \hat{r_i}$,} where $\hat{r_i}$ is 
 a unit vector in the transverse plane pointing to the center of 
 calorimeter tower $i$ and $\Et^i$ is the transverse
 energy deposited in that tower. Only towers with $|\eta|<$ 3.6 are included.
\end{thebibliography}
\end{document}